\documentclass[a4paper,11pt]{karticle}
\usepackage{graphicx}

\renewcommand{\thesection}{\S\arabic{section}}

\title
{
Generalized Amplitude Truncation of Gaussian $1/f^{\alpha}$ noise}

\author
{ 
Donghak {\sc Choi}
and Nobuko {\sc Fuchikami}
}

\date
{(\today)
}

\begin{document}
\sloppy
\maketitle
\begin{center}
\textit{
Department of Physics, Tokyo Metropolitan University, 
Minami-Ohsawa, Hachioji, Tokyo 192-0397, Japan }
\end{center}

\begin{abstract}
\setlength{\baselineskip}{15pt}
We study a kind of filtering, 
an amplitude truncation with 
upper and lower truncation levels $x_{\rm{max}}$ and 
$x_{\rm{\min}}$. This is a generalization of 
the simple transformation 
$y(t)=\rm{sgn}[\it{x(t)}]$, for which a rigorous 
result was obtained recently. So far numerical experiments 
have shown that a power law spectrum $1/f^{\alpha}$ seems to be 
transformed again into a power law spectrum $1/f^{\beta}$ under 
rather general condition for the truncation levels. 
We examine the above numerical results analytically. 
When $1<\alpha<2$ and 
$x_{\rm{max}}=-x_{\rm{min}}=a$, 
the transformed spectrum is shown to be 
characterized by 
a certain corner frequency $f_{\rm c}$ 
which divides the spectrum into two parts with different exponents.  
We derive $f_{\rm c}$ depending on $a$ as $f_{\rm c} \sim a^{-2/(\alpha-1)}$. 
It turns out that the output signal should deviate 
from the power law spectrum when the truncation is asymmetrical. 
We present a numerical example such that 
$1/f^2$ noise converges to $1/f$ noise 
by applying the transformation $y(t)=\rm{sgn}[\it{x(t)}]$ repeatedly. 
\end{abstract}

{\footnotesize
\hspace{0.3cm}\textsf{KEYWORDS:}~\textsf{1/f fluctuation, nonlinear transformation, filtering, power law}
}

\newcommand{\lsim}{\begin{minipage}{12pt}
\vspace{0pt}$
\,\stackrel{\textstyle <}{\sim}$\end{minipage}}
\newcommand{\gsim}{\begin{minipage}{12pt}
\vspace{0pt}$
\,\stackrel{\textstyle >}{\sim}$\end{minipage}}
\baselineskip 4.5mm

\section{Introduction}
\setlength{\baselineskip}{17pt}
$1/f$ noise whose power spectrum is inversely proportional to 
the frequency $f$ has been observed in a variety of systems, 
since it was first discovered in current fluctuations of a vacuum 
tube (see ref. 1 and references therein). 
Sometimes  $1/f^{\alpha}$ noise with the exponent $\alpha$ not close 
to $1$ but between $0$ and $2$ is also referred to as 
`$1/f$ noise'\cite{bak97,frieden94}. 
When this extended definition is applied, `$1/f$ noise' is more 
widely observed\cite{rf:mandel}. 

\begin{figure}[htbp]
\vspace{0.4cm}
	\begin{center}
	\includegraphics[width=0.82\linewidth]{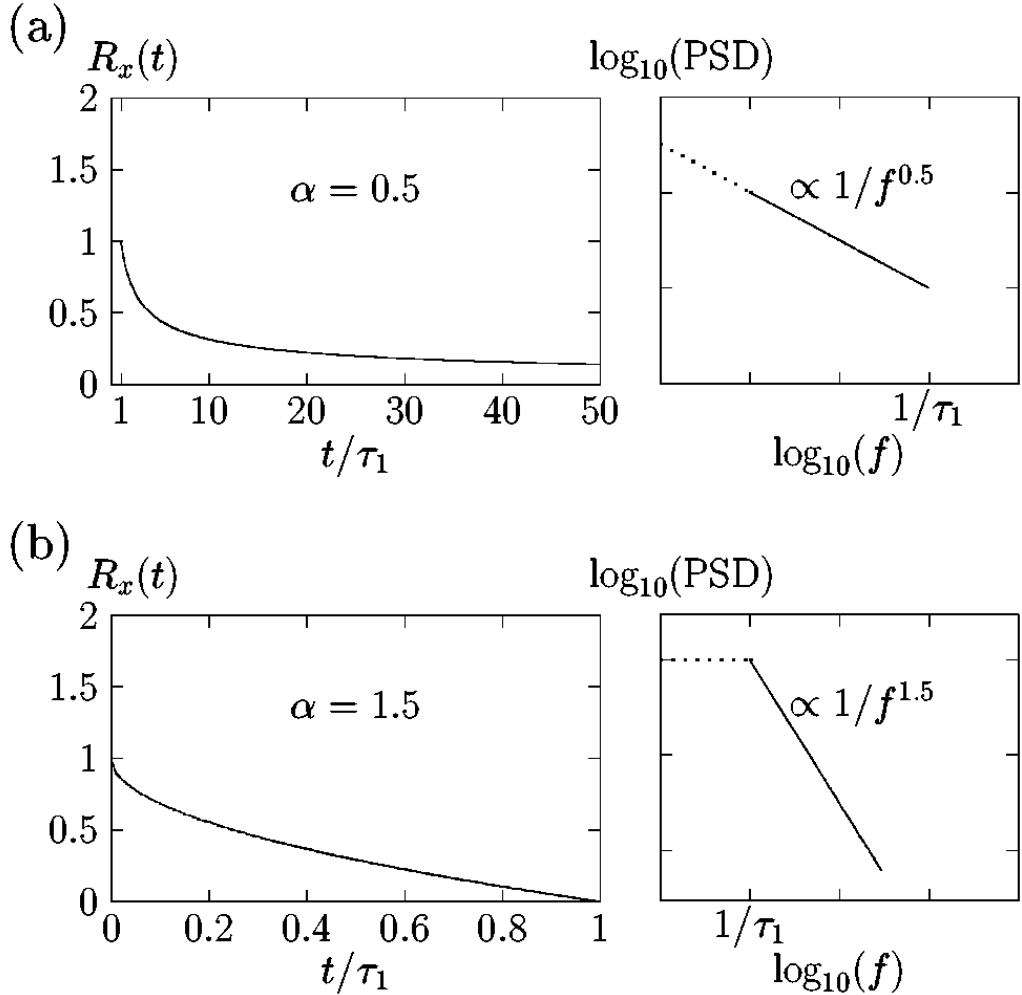}
	\end{center}
	\caption{Correlation function $R_x(t)$ of a $1/f^{\alpha}$ noise.  
%yielding power law spectrum $1/f^{\alpha}$.
(a) $0< \alpha <1$: eqs. (\ref{eqn:incor1}). 
(b) $1< \alpha <2$: eqs. (\ref{eqn:incor222}).}
\label{fig:1}
\end{figure}

It should be noted that the integral $\int_0^{\infty}f^{-\alpha}df$ 
diverges for all $\alpha$. Therefore, if the process is stationary, 
which is usually a reasonable assumption, a low or high frequency 
cut off should exist corresponding to $\alpha \geq 1$ or 
$0 \leq \alpha \leq1$, respectively. Or equivalently, the value of 
$\alpha$ cannot be constant: $\alpha$ should become smaller 
(larger) than unity in a low (high) frequency range. 
The power law behavior involves several difficulties. 
The autocorrelation function of $1/f^{\alpha}$ 
noise with $\alpha < 1$ decays very slowly (power law decay, 
see eq. (\ref{eqn:incor1b}) and Fig. \ref{fig:1}(a)) 
especially when $\alpha$ is close to unity, unless there is another low 
frequency cut off $f_0$ so that the spectrum becomes white ($\alpha=0$) 
in a range $f\ll f_0$. It often happens that the cut off frequency 
$f_0$ is extremely low or not observed during the practical observation 
time.\cite{rf:mandel,musha90} One of the difficulties is 
to explain a long time scale $\sim 1/f_0$ 
from any realistic model. Another problem is the value of the exponent which 
largely deviates from $2$. The exponent $\alpha=2$ is most naturally 
expected in nonstationary processes. Random walk in 
$n$-dimensional space yields $1/f^2$ spectrum. 
Also, a Lorentzian spectrum 
$(f^2+f_0^2)^{-1}$ looks as $1/f^2$ when the observation time is not long 
enough: $f\gg f_0$. Lorentzian spectrums are obtained from Debye-type 
relaxation processes, which occur most commonly. A randomly amplified 
Langevin system which has a power law distribution function also leads 
to a Lorentzian spectrum,\cite{fuchi99} 
even though the process is nonstationary so that 
the Wiener-Khintchine theorem does not hold.  
Then what mechanism could exist that would reduce the value of 
$\alpha$ from $2$? The present work relates to this second problem. 
We shall investigate, by generalizing a previous 
theory\cite{ishioka00}, how Gaussian $1/f^{\alpha}$ noise is affected by a 
kind of filtering, an amplitude truncation. 

In the previous paper, a simple dichotomous transformation defined by 
$y(t)=\rm{sgn}[\it{x(t)}]$ 
 for Gaussian 
$1/f^{\alpha}$ noise 
was studied in detail and  a rigorous 
result for transformation properties 
was obtained\cite{ishioka00}. 
When a Gaussian noise $x(t)$ with a power law spectrum $1/f^{\alpha}$ 
is filtered by this transformation, power spectral density 
(PSD) of the output noise $y(t)$ obeys 
again a power law $1/f^{\beta}$. 
The exponent $\beta$ is derived as $\beta=\alpha$ for 
$0<\alpha\leq1$, and $\beta=(\alpha+1)/2$ for $1\leq \alpha<2$. 

The present paper deals with a more general amplitude truncation 
\begin{equation}
y(t)=\left\{
\begin{array}{ll}
 x_{\rm{min}}  &\mbox{ if }\;\;\; x(t) \leq x_{\rm{min}}\mbox{, }  \\
 x_{\rm{max}}  &\mbox{ if }\;\;\; x(t) \geq x_{\rm{max}}\mbox{, }  \\
 x(t)   & \mbox{ otherwise, }
% x(t)   & \mbox{ if }\;\;\; x_{\rm{min}} < x(t) < x_{\rm{max}} \mbox{ , }
\end{array}
\right.
\label{eqn:truncation}
\end{equation}
for Gaussian $1/f^{\alpha}$ noise. 
%$1/f^{\alpha}$ noise $x(t)$. 
The transformation $y(t)=\rm{sgn}[\it{x(t)}]$ 
is equivalent to the situation $x_{\rm{max}}=-x_{\rm{min}}\ll1$ 
in eq. (\ref{eqn:truncation}), namely, the dichotomous 
transformation is a special case of 
eq. (\ref{eqn:truncation}).
So far numerical experiments have shown\cite{gingl96,gingl99} 
that a power law spectrum $1/f^{\alpha}$ seems to be 
transformed again into a power law spectrum $1/f^{\beta}$ under 
rather general condition for the truncation levels
$x_{\rm{min}}$ and $x_{\rm{max}}$. 
However, the numerical results are not always clear, so that 
we will investigate analytically the above amplitude truncation to 
know how $\beta$ depends on the levels as well as $\alpha$ 
and when the output signal deviates from the power law spectrum.

The rest of the present paper is organized as follows. In the next 
section, we briefly review and reinterpret 
the recent result for the dichotomous transformation. 
In \S 3, we extend our study for the 
symmetrical truncation with finite levels
: $x_{\rm{max}}=-x_{\rm{min}}=a$ 
in (\ref{eqn:truncation}). 
We consider the asymmetrical truncation in \S 4. 
It will be shown how PSD of the output signal deviates from the 
power law. 
In \S 5, we present a numerical example such that 
$1/f^2$ noise converges to $1/f$ noise 
by applying the dichotomous 
transformation repeatedly. 
The last section is devoted to summary and discussions. 

\section{Dichotomous Transformation}
\setcounter{equation}{0}
The dichotomous 
transformation is defined by 
\begin{eqnarray}
y(t)=  \rm{ sgn }
[\mit x(t)]= \left\{
\begin{array}{ll}
 -1  &\mbox{ if }\;\;\; x(t) \leq 0 \mbox{ , } \\
 +1  &\mbox{ if }\;\;\; x(t) > 0 \mbox{ , }  \\
\end{array}
\right.
\label{eqn:dichotomous}
\end{eqnarray}
where $x(t)$ is an input noise with zero mean and 
$y(t)$ is an output noise. This is a special case of 
the amplitude truncation (\ref{eqn:truncation}) 
and was investigated in detail in ref. 1.
It turned out that when the transformation (\ref{eqn:dichotomous}) 
is applied to a Gaussian $1/f^{\alpha}$ 
noise $x(t)$, then the output noise $y(t)$ exhibits a power law spectrum 
$1/f^{\beta}$. 
The exponent $\beta$ was derived as 
$\beta=\alpha$ for $0<\alpha\leq 1$, 
and $\beta=(\alpha+1)/2$ for $1\leq \alpha <2$. 
As will be shown below, a key to understand this result is a 
transformation property of the correlation function 
$R(t)$ (eq. (\ref{eqn:relation})). 
%which also obeys a power law 
%$t^{\alpha-1}$ when PSD obeys the power law $f^{-\alpha}$.

The correlation function of the 
output signal $y(t)$ is obtained from eq. (\ref{eqn:dichotomous}) as 
\begin{eqnarray}
R_{y}(t)& \equiv & \langle y(t_0)y(t_0+t) \rangle \nonumber \\        
        & = &1 \cdot P\left( y(0)y(t)=1 \right)+(-1) 
\cdot P\left(y(0)y(t)=-1\right) \nonumber \\
        & = &P\left(x(0)x(t)>0\right)-P\left(x(0)x(t)<0\right) \nonumber \\
        & = &2 \cdot P\left(x(0)x(t)>0\right)-1 \mbox{ , }
\label{eqn:outcorre}
\end{eqnarray}
where $P(.)$ stands for the probability that the condition of the 
argument is satisfied. 
In the above, we have assumed a stationary process for the input noise, 
which leads to a stationary process for the output one. 
We can therefore apply the Wiener-Khintchine theorem to both input and 
output noises. 
We also assume that the process for the input noise is correlated 
Gaussian\cite{rf:note}. 
That is, the joint probability is given by 
\begin{eqnarray}
P(x(t_{0}),x(t_{0}+t))& = &P(x(0),x(t)) \nonumber \\
                   & = &\frac{1}{A_{0}}\exp[-(x^{2}-2cxy+y^{2})/B] 
\nonumber \\
                     & \equiv &f(x,y,c) \mbox{ , }
\label{eqn:joint}
\end{eqnarray}
where,
%$x \equiv x(0)$, $ y \equiv x(t_{1}-t_{2})$, $c \equiv R_{x}(t_{1}-t_{2})$,
% $A_{0} \equiv 2\pi R_{x}(0)\sqrt{1-c^{2}}$, $B \equiv 2R_{x}(0)(1-c^{2})$, 
%$R_{x}(t)$ being the correlation function of $x(t)$. 
\begin{equation}
\left\{
\begin{array}{ll}
x \equiv x(0)\mbox{ , }  \\
y \equiv x(t)\mbox{ , } \\
c \equiv R_{x}(t)\mbox{ , } \\
A_{0} \equiv 2\pi R_{x}(0)\sqrt{1-c^{2}}\mbox{ , } \\
B \equiv 2R_{x}(0)(1-c^{2})\mbox{ , } \\
\end{array}
\right.
\label{eqn:definjoint}
\end{equation}
and 
\begin{eqnarray}
R_{x}(t) \equiv  \langle x(t_0)x(t_0+t) \rangle 
\label{eqn:definjoint2}
\end{eqnarray}
is the correlation function of $x(t)$.

The probability $P(xy>0)$ is calculated from eq. (\ref{eqn:joint}) as 
\begin{eqnarray}
P(xy>0)&=&\int_{0}^{\infty}dx\int_{0}^{\infty}dy f(x,y,c) \nonumber \\
&&+\int_{-\infty}^{0}dx\int_{-\infty}^{0}dy f(x,y,c) \nonumber \\
&=&\frac{1}{2}+\frac{1}{\pi}\arcsin[R_{x}(t)] \mbox{ . }
\end{eqnarray}

Thus we obtain the relation between the input and output 
correlation functions as 
\begin{eqnarray}
R_{y}(t)=\frac{2}{\pi}\arcsin[R_{x}(t)] \label{eqn:relation} \mbox{ . }
\end{eqnarray}
%The relation is correct if input noise is Gaussian.
%The relation (\ref{eqn:relation}) 
%leads to the relation between PSD 
%by using the Winer-Khinchin theorem. 

The previous result for the exponent $\beta$ can easily be understood 
qualitatively if we note the approximated expression for 
(\ref{eqn:relation}): 
%\begin{equation}
%\left\{
%\begin{array}{ll}
%\mbox{ when } |R_x(t)| \ll 1 \;\;\;\;\;\;\;\; 
%R_y(t)\propto R_x(t) \label{eqn:sssss1}\\
%\mbox{ when } |1-R_x(t)| \ll 1 \;\; R_y(t)\propto \mbox{const} 
%-(1-R_x(t))^{\frac{1}{2}}
%\label{eqn:sssss2}
%\end{array}
%\right.
%\end{equation}
\begin{eqnarray}
&&\mbox{ when } |R_x(t)| \ll 1 \mbox { , }\;\;\;\;\;\;\;\; 
R_y(t)\propto R_x(t); \label{eqn:sssss1}\\
&&\mbox{ when } |1-R_x(t)| \ll 1 \mbox{ , } \;\; R_y(t)\propto \mbox{const.} 
-(1-R_x(t))^{\frac{1}{2}} \mbox{ . }
\label{eqn:sssss2}
\end{eqnarray}

%%%%%%%%%%%%%%% fig. 1
%%%%%%%%%%%%%%%

First we explain $\beta=\alpha$ for $0<\alpha<1$. 
When PSD for low frequencies obeys the power law as $\sim f^{-\alpha}$ 
with $0< \alpha <1$, the corresponding correlation function $R_x(t)$ 
should be $R_x(t)\sim t^{\alpha-1}$ for large $t$ values\cite{ishioka00,stanly94}. 
Let us consider a system whose characteristic time scale is $\tau_1$. 
Then, by assuming $R_x(t)$ as 
%\begin{equation}
%R_x(t)=\left\{
%\begin{array}{ll}
% 1 \;\;\;\;\;\;\;\;\;\;\;\;\; t/\tau_1 \leq 1 \mbox{ , }  \\
% (t/\tau_1)^{\alpha-1}\;\;\; t/\tau_1 \geq 1\mbox{ , }  
%\end{array}
%\right.
%\label{eqn:incor1}
%\end{equation}
\begin{subeqnarray}
\label{eqn:incor1}
R_x(t)&=& 1 \;\;\;\;\;\;\;\;\;\;\;\;\;\;\;\; 
\mbox{ for }\;\;\;t/\tau_1 \leq 1 \mbox{ , }\\
\slabel{eqn:incor1b}
R_x(t)&=& (t/\tau_1)^{\alpha-1} \;\;\; 
\mbox{ for }\;\;\;t/\tau_1 \geq 1 \mbox{ , }  
%R_x(t)&=& 1 \mbox{ , }\;\;\;\;\;\;\;\;\;\;\;\;\;\;\;\; t/\tau_1 \leq 1 
%\mbox{ , }  \\
%\slabel{eqn:incor1b}
%R_x(t)&=& (t/\tau_1)^{\alpha-1}\mbox{ , }\;\;\; t/\tau_1 > 1\mbox{ , }  
\end{subeqnarray}
the PSD is $\sim (\tau_1 f)^{-\alpha}$ for low frequencies: 
$0<f \ll 1/\tau_1$. 
The relation  (\ref{eqn:sssss1}) can be used to obtain $R_y(t)$ for 
large $t$: $t/\tau_1\gg1$, because $R_x(t)$ is small. This is why the same 
exponent $\beta=\alpha$ is obtained for $0<\alpha<1$. 

The case $1<\alpha<2$ can also be understood qualitatively. Let us 
assume the correlation function as 
%\begin{equation}
%R_x(t)=\left\{
%\begin{array}{ll}
% 1-(t/\tau)^{\alpha-1} \;\;\; t/\tau_1 \leq 1 \mbox{ , }  \\
% 0 \;\;\;\;\;\;\;\;\;\;\;\;\;\;\;\;\;\;\;\;\; t/\tau_1 \geq 1\mbox{ . }  
%\end{array}
%\right.
%\label{eqn:incor2}
%\end{equation}
\begin{subeqnarray}
\label{eqn:incor222}
\slabel{eqn:incor2}
R_x(t)&=& 1-(t/\tau_1)^{\alpha-1} \;\;\; 
\mbox{ for }\;\;\; t/\tau_1 \leq 1 \mbox{ , }  \\
\slabel{eqn:incorb}
R_x(t)&=& 0 \;\;\;\;\;\;\;\;\;\;\;\;\;\;\;\;\;\;\;\;\;\; 
\mbox{ for }\;\;\; t/\tau_1 \geq 1\mbox{ . }  
\end{subeqnarray}
The above correlation function yields a white spectrum if the observation 
time $t$ is long enough: $t\gg\tau_1$, i.e., $f\ll1/\tau_1$. 
Naturally we are interested in the situation in which the characteristic 
time scale $\tau_1$ is very long. The power law, eq. 
(\ref{eqn:incor2}) again leads to the power law spectrum 
$S_x(f)\propto(\tau_1f)^{-\alpha}$ for a wide range: $f\gg1/\tau_1$\cite{ishioka00}.
Note that if $\tau_1$ is very long, the lower limit $1/\tau_1$ can be 
small and the power law $\sim f^{-\alpha}$ holds practically in a wide 
range of low frequencies. 
When time $t$ is long enough but satisfies $t\ll\tau_1$, the 
approximation (\ref{eqn:sssss2}) can be used for $R_y(t)$: 
\begin{eqnarray}
R_y(t) &\propto& \mbox{ const.}-(t/\tau_1)^{\frac{\alpha-1}{2}} \nonumber \\
&\equiv& \mbox{ const.}-(t/\tau_1)^{\beta-1}\mbox{ . }
\label{eqn:sec2las}
\end{eqnarray}
Corresponding to the above $R_y(t)$, the PSD is obtained as $\sim f^{-\beta}$, 
where $\beta=(\alpha+1)/2$.

\section{Symmetrical Truncation with Finite Levels}
\setcounter{equation}{0}

In this section we generalize the results obtained 
in the above.
Suppose that a Gaussian $1/f^{\alpha}$ noise with zero mean is 
transformed by a symmetrical truncation with finite levels 
$x_{\rm{max}}=-x_{\rm{min}}\equiv a$. 
Let us define a typical time scale $\tau_{\rm c}$ in which the noise signal 
passes between the two truncation levels. 
In a high frequency range $f\gg f_{\rm c}\equiv 1/\tau_{\rm c}$, PSD of 
the truncated noise is mainly determined by the behavior of the noise 
between the two levels. 
On the other hand, in a low frequency range 
$f \stackrel{\displaystyle <}{\raisebox{-1ex}{$\sim$}}  f_{\rm c}$ 
(corresponding 
to a long-term observation), PSD of the output signal should have a 
similar PSD to that obtained by the dichotomous transformation. 
Therefore, for $0<\alpha<1$, the same exponent $\beta=\alpha$ is 
expected both in high 
($f_{\rm c}\stackrel{\displaystyle <}{\raisebox{-1ex}{$\sim$}} f \ll 1/\tau_1$) 
and low ($f\stackrel{\displaystyle <}{\raisebox{-1ex}{$\sim$}} f_{\rm c}$) 
frequency ranges. 

\begin{figure}[htbp]
\vspace{0.4cm}
	\begin{center}
	\includegraphics[width=0.38\linewidth]{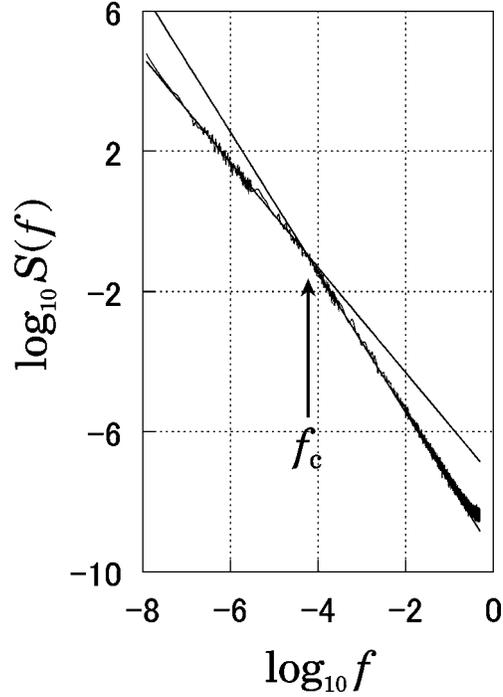}
	\end{center}
	\caption{PSD of the truncated signal for $\alpha=2$. 
The PSD is obtained by averaging over $40$ samples. PSDs in three 
frequency ranges are composed. Straight lines are $S(f) \propto f^{-2}$ 
and $\propto f^{-1.5}$}
\label{fig:2}
\end{figure}

In contrast, when $1<\alpha<2$, PSD is divided into two parts by a 
corner frequency $f_{\rm c}$: $\beta=\alpha$ for a high frequency range 
($f\stackrel{\displaystyle >}{\raisebox{-1ex}{$\sim$}} f_{\rm c}$) 
and $\beta=(\alpha+1)/2$ for a low frequency range 
($1/\tau_1\ll f \stackrel{\displaystyle <}{\raisebox{-1ex}{$\sim$}} f_{\rm c}$). 
A typical PSD of a truncated signal for $\alpha=2$ 
is shown in Fig.~\ref{fig:2}.
%%%%%%%%%%%%%%% fig. 2
%%%%%%%%%%%%%%%

The aim in this section is 
to derive the dependence of $f_{\rm c}$ upon the truncation level $a$. 
The transformation is expressed as 
\begin{equation}
y(t)=\left\{
\begin{array}{ll}
 x(t)                                &\mbox{ if } 
\;-a \leq x(t) \leq a \mbox{, } \\
 a \cdot \rm{ sgn } [\mit x(t)]      &\mbox{ otherwise, }
\end{array}
\right.
\label{eqn:astranin}
\end{equation}
where $x(t)$ is a correlated Gaussian noise with zero mean.

As in the previous section, the joint probability (\ref{eqn:joint}) yields
the correlation function 
of the output signal as 
\begin{eqnarray}
R_{y}(t)& = & 2 \cdot a^2 (I(c)-I(-c))+ 
2 \cdot (J(c)-J(-c)) \nonumber \\
        && + 4 \cdot a (K(c)-K(-c)) \mbox{ , }
\label{eqn:tranlevco}
\end{eqnarray}
where
\begin{eqnarray}
I(c)\equiv\int_{a}^{\infty} dx\int_{a}^{\infty} dy 
f(x,y,c)\mbox{ , } \nonumber \\
J(c)\equiv\int_{0}^{a}dx\int_{0}^{a}dy\;xyf(x,y,c) \mbox{ , }      \nonumber \\
K(c)\equiv\int_{a}^{\infty}dx \int_{0}^{a}dy\; yf(x,y,c)\mbox{ . } \nonumber 
%f(x,y,c)\equiv\frac{1}{A_0}\exp [-(x^2-2cxy+y^2)/B]. \nonumber
\end{eqnarray}
If we assume $a\ll 1$, 
$I(c)$, $J(c)$ and $K(c)$ can be expanded with respect to $a$ as 
\begin{eqnarray}
I(c)&=&\frac{1}{4}[1+\frac{2}{\pi}\arcsin (c)]
      -a\frac{\sqrt{B\pi}}{A_0}
      -a^2\frac{c-1}{A_0}+O(a^3) \mbox{ , } \nonumber \\
J(c)&=&a^4\frac{6}{4!A_0}+O(a^5) \mbox{ , }\nonumber \\
K(c)&=&a^2\frac{\sqrt{B\pi}}{2!2A_0}+a^3\frac{2c-3}{3!A_0}
+O(a^4) \mbox{ . }
\label{eqn:ijk}
\end{eqnarray}
Substitution of Eq. (\ref{eqn:ijk}) into (\ref{eqn:tranlevco}) 
leads to the correlation function of the output signal as 
\begin{eqnarray}
R_y(t)
        &=&2a^2\{\frac{1}{\pi}\arcsin(c)
-2\cdot\frac{ca^2}{2\pi \sqrt{1-c^2}}+\cdots\} \nonumber \\
        && +4a\{\frac{ca^3}{3\pi\sqrt{1-c^2}}+\cdots\} \nonumber \\
        & = & a^2\{\frac{2}{\pi}\arcsin(c)-\frac{2ca^2}{\pi\sqrt{1-c^2}}
+\frac{4ca^2}{3\pi\sqrt{1-c^2}}+\cdots \}  \mbox{ . }\nonumber
%4a^2\frac{c}{\pi\sqrt{1-c^2}} \nonumber \\
%        && \; \; \; \; \; \; \; +a\sqrt{\frac{2}{\pi}}-a^2\frac{2}
%{\pi\sqrt{1-c^2}}\cdots\} \nonumber
\label{eqn:perturbation}
\end{eqnarray}
Since we are considering the case of $1<\alpha<2$ , we substitute 
$c\equiv R_x(t)=1-t^{\alpha-1}$ (eq. (\ref{eqn:incor2})) 
into the above.  The correlation 
function is thus obtained as 
\begin{eqnarray}
R_y(t) 
                         &\simeq& a^2\{1-\frac{2\sqrt{2}}{\pi}
t^{(\alpha-1)/2}\nonumber \\
    &&-a^2\frac{\sqrt{2}}
{3\pi}t^{-\frac{\alpha-1}{2}}+\cdots \}, 
\label{eqn:pertur_ts}
\end{eqnarray}
for $t\ll 1 $.  
Here and hereafter, the dimensionless time $t/\tau_1$ is replaced 
by $t$ for simplicity. Thus $t\ll1$ actually means $t\ll\tau_1$. 
If $a$ is small enough and/or $t$ is not so small so that 
the third term in the braces in (\ref{eqn:pertur_ts}) can be neglected, 
$R_y(t)$ reduces to eq. (\ref{eqn:sec2las}). 
As $t$ decreases, the third term grows and becomes comparable to the 
second term. In other words, the perturbative expansion 
(\ref{eqn:pertur_ts}) breaks down when 
$t^{(\alpha-1)/2}\sim a^2 t^{-(\alpha-1)/2}$ i.e. $t\sim a^{2/(\alpha-1)}$. 
This means that the corner frequency $f_{\rm c}$ depends on the truncation 
level $a$ as 
\begin{eqnarray}
f_{\rm c} \sim a^{-2/(\alpha-1)}.  
\label{eqn:dependence}
\end{eqnarray}  
For a long-term observation, i.e., 
$1 \gg t\stackrel{\displaystyle >}{\raisebox{-1ex}{$\sim$}} 
a^{2/(\alpha-1)}$ 
(which corresponds to 
$1 \ll f\stackrel{\displaystyle <}{\raisebox{-1ex}{$\sim$}} f_{\rm c}$), 
the exponent $\beta$ of the PSD becomes 
$\beta=(\alpha+1)/2$ because (\ref{eqn:pertur_ts}) reduces to 
(\ref{eqn:sec2las}). 
For a high frequency range: 
$f\stackrel{\displaystyle >}{\raisebox{-1ex}{$\sim$}} f_{\rm c}$, 
the same exponent 
$\beta=\alpha$ as the input signal 
is expected by the reason mentioned already, 
although the correlation function of the form 
$R_y(t)\sim \mbox{const.}-t^{\alpha -1}$ (which corresponds to 
the PSD$\sim f^{-\alpha}$) cannot be obtained from any correction 
terms added to (\ref{eqn:pertur_ts}) because the perturbative expansion 
has broken down. One can see in 
Fig.~\ref{fig:2} that the PSD obtained from the 
numerical 
simulation is really composed of two power law spectra separated by 
the corner frequency $f_{\rm c}$. 

The scaling property of eq. (\ref{eqn:dependence}) can also be suggested 
from the following self-affine character of the signal from the fractional 
Brownian motion:\cite{vicsek92} 
\begin{eqnarray}
\langle x(t)^2 \rangle \sim t^{2H} .
\label{eqn:selfaffin}
\end{eqnarray}
It is well known that if a signal satisfies 
the scaling relation 
(\ref{eqn:selfaffin}), 
then its PSD is  $1/f^{\alpha}$  with $\alpha=2H+1$ where 
$\alpha\geq 0$\cite{vicsek92}.
The above self-affine character yields  
\begin{eqnarray}
\sqrt{\langle x(t)^2 \rangle} 
&=& 
\sqrt{\langle x(a^{-\frac{1}{H}}a^{\frac{1}{H}}t)^2 \rangle} \nonumber \\
&\propto& \frac{1}{a}\sqrt{\langle x(a^{\frac{1}{H}}t)^2 \rangle}.
\end{eqnarray}
This means that if the truncation levels are changed 
from $ \pm U$ to $ \pm aU$ and the time 
%scale 
is changed from $t$ to $a^{1/H}t$ simultaneously, the structure 
of the noise is invariant. Then the corner 
frequency changes from $f_{\rm c}$ to $a^{-(1/H)}f_{\rm c}=a^{-2/(\alpha-1)}f_{\rm c}$. 
This coincides with the result obtained from the 
perturbation method; the present 
argument can be applicable even for large $a$ values. 

\begin{figure}[htbp]
\vspace{0.4cm}
	\begin{center}
	\includegraphics[width=0.55\linewidth]{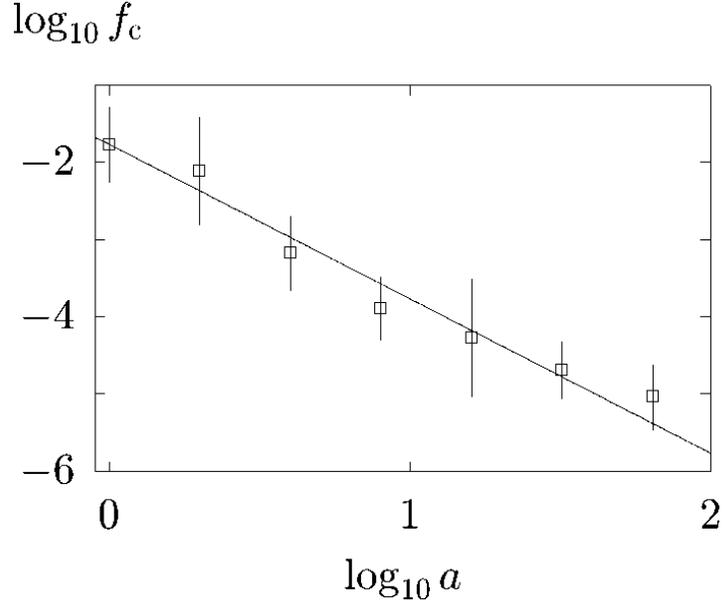}
	\end{center}
	\caption{Log-log plot of the corner frequency $f_{\rm c}$ vs the truncation level $a$ 
for $\alpha=2$. The straight line is $f_{\rm c}\propto a^{-2}$.}
\label{fig:3}
\end{figure}

To confirm $a$-dependence of $f_{\rm c}$, 
eq. (\ref{eqn:dependence}), we performed a numerical simulation 
for $\alpha=2$ by varying the truncation level $a$ in eq. (\ref{eqn:astranin}). 
For each $a$ value, a PSD $S(f)$ was obtained by averaging over $40$ samples. 
Then $\log S(f)$ was fitted to the line 
%$F(f)=A(f_{\rm c}^{0.5}f^{-2}\Theta (f-f_{\rm c})+f^{-1.5}\Theta (f_{\rm c}-f)$ so as to 
$F(f)=(-1.5\log f+1.5\log f_{\rm c}+A)\Theta(\log f_{\rm c}-\log f)
+(-2.0\log f+2.0\log f_{\rm c}+A)\Theta(\log f-\log f_{\rm c})$ 
where 
$\Theta(x)$ is a Heaviside function with $\Theta(0)=1/2$. 
The result is shown in Fig.~\ref{fig:3}. 
Agreement between the theoretical line: $f_{\rm c}\propto a^{-2}$ and the numerical 
result is tolerable. 
%%%%%%%%%%%%%%% fig. 3
%%%%%%%%%%%%%%%

\section{Asymmetrical Truncation}
\setcounter{equation}{0}

We consider the simplest case, i.e., the following asymmetrical 
dichotomous transformation: 
\begin{equation}
y(t)=\rm{ sgn } [\mit x(t)-a]=\left\{
\begin{array}{ll}
 +1            & \mbox{ if }\;\;\; x(t) \geq a  \mbox{, } \\
 -1            & \mbox{ if }\;\;\; x(t) < a  \mbox{. }
\end{array}
\right.
\label{eqn:asymtrans}
\end{equation}
The input signal is a Gaussian $1/f^{\alpha}$ noise with 
zero mean, and the condition 
$0<\alpha<1$ is assumed here. 

\begin{figure}[htbp]
\vspace{0.4cm}
	\begin{center}
	\includegraphics[width=0.50\linewidth]{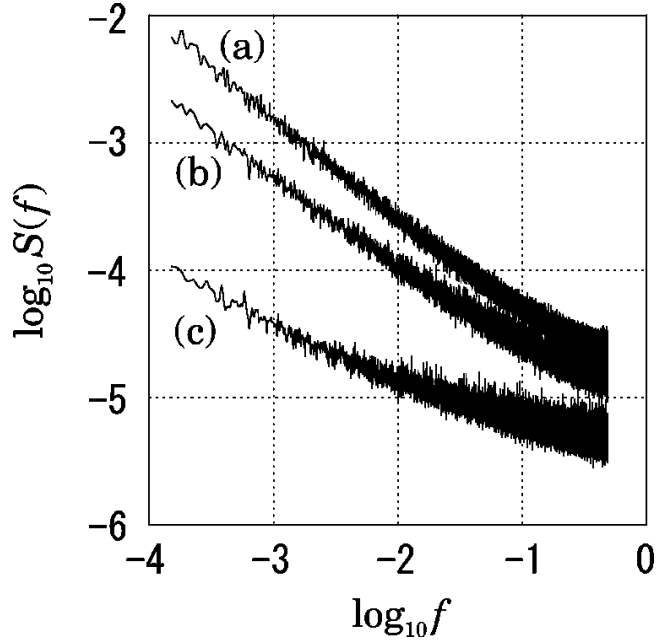}
	\end{center}
	\caption{Output PSDs obtained from the asymmetric transformation 
(\ref{eqn:asymtrans}) where the exponent of the input noise is $\alpha=0.8$. 
Each PSD was obtained by averaging over 40 samples.
(a), (b), and (c) correspond to the truncation level 
$a=0$, $\sigma$ and 2$\sigma$, 
respectively, where $\sigma$ is the standard deviation of the input signal. 
}
\label{fig:4}
\end{figure}

Figure ~\ref{fig:4} illustrates the PSD of the output signal obtained 
by eq. (\ref{eqn:asymtrans}) for $\alpha =0.8$ and 
$a=0, \sigma, 2\sigma$, where $\sigma$ is the 
standard deviation of the input signal. 
It is unclear from these numerical results 
whether the PSD obeys the power law or not. 
%To make certain the question we investigate the transformation 
%(\ref{eqn:asymtrans}) analytically.
%%%%%%%%%%%%%%% fig. 4
%%%%%%%%%%%%%%%

Note that the mean value of the output signal is finite: 
\begin{eqnarray}
\langle y(t) \rangle &=& \int_{-\infty}^{\infty}\mbox{sgn}[x-a]f(x)dx \nonumber \\
&=&-2\int_{0}^{a}f(x)dx \equiv \bar{y} \mbox{ , }
\label{eqn:secfortw}
\end{eqnarray}
where $f(x)$ is the Gaussian distribution function for the input signal: 
\begin{eqnarray}
f(x)=\frac{1}{\sqrt{2\pi R_x(0)}}\exp\left[-\frac{x^2}{2R_x(0)}\right] \mbox{ . }
\end{eqnarray}
The correlation function of $y(t)$ is thus given by 
\begin{eqnarray}
R_y(t)&=&\langle (y(t_0)-\bar{y})(y(t_0+t)-\bar{y} \rangle \nonumber \\
      &=&\langle y(t_0)y(t_0+t)\rangle-(\bar{y})^2 \\
      &=&2P((x(0)-a)(x(t)-a)>0)-1-(\bar{y})^2 \mbox{ . }
\label{eqn:realascor}
\end{eqnarray}
The probability $P((x(0)-a)(x(t)-a)>0)$ is expressed as 
\begin{eqnarray}
&&P((x(0)-a)(x(t)-a)>0) \nonumber \\
&=&\bigm[2\int_{-\infty}^{\infty}dx\int_{0}^{a}dy
+2\int_{a}^{\infty}dx\int_{a}^{\infty}dy\bigm]
f(x,y,c) \mbox{ , }
\label{eqn:asymcorr}
\end{eqnarray}
where $f(x,y,c)$ is the joint probability defined by 
(\ref{eqn:joint}) $\sim$ (\ref{eqn:definjoint2}). 
The first term in (\ref{eqn:asymcorr}) depends on $a$ but not on $t$: 
\begin{eqnarray}
2\int_{-\infty}^{\infty}dx\int_{0}^{a}dyf(x,y,c)
=2\int_{0}^{a}dxf(x)dx=-\bar{y} \mbox{ , }
\end{eqnarray}
which yields 
\begin{eqnarray}
P((x(0)-a)(x(t)-a)>0)=-\bar{y}+T \mbox { , }\label{eqn:pasy}\\
T\equiv 2\int_{a}^{\infty}dx\int_{a}^{\infty}dyf(x,y,c) \mbox{ . }
\end{eqnarray}
We calculate $R_y(t)$ in two cases $a\ll 1$ and $a\gg1$. 
Since $R_x(t)$ is assumed as eqs. (\ref{eqn:incor1}), 
we set $R_x(0)=1$ hereafter. 

When $a\ll 1$, $\bar{y}$ is expanded with respect to $a$: 
\begin{eqnarray}
\bar{y}=-\frac{2a}{\sqrt{2\pi}}+O(a^3) \mbox{ . }
\label{eqn:outmean}
\end{eqnarray}
The second term $T$ in (\ref{eqn:pasy}) is similarly expanded as 
\begin{eqnarray}
T\equiv \frac{1}{2}+\frac{1}{\pi}\arcsin[c]-\frac{2}{\sqrt{2\pi}}a
+\frac{1-c}{\pi\sqrt{1-c^2}}a^2+\cdots \mbox{ . }
\end{eqnarray}
For $t\gg1$ the correlation function becomes $c\equiv R_x(t)=t^{\alpha-1}\ll 1$. 
Then $T$ can be approximated as 
\begin{eqnarray}
T=\frac{1}{2}-\frac{2}{\sqrt{2\pi}}a+\frac{1}{\pi}c
+\frac{1}{\pi}(1-c+\frac{1}{2}{c^2})a^2+O(c^3) \mbox{ . }
\label{eqn:therdterm}
\end{eqnarray}
Substituting  (\ref{eqn:pasy}), (\ref{eqn:outmean}), 
(\ref{eqn:therdterm}) and $c\equiv 
t^{\alpha-1}$ into (\ref{eqn:realascor}), we obtain 
\begin{eqnarray}
\label{eqn:asasco1}
R_y(t)
&=&2(-\bar{y}+T)-1-\bar{y}^2 \\
&\sim& \frac{2}{\pi}(1-a^2)c+\frac{a^2}{\pi}c^2 \nonumber \\
&=& \frac{2}{\pi}(1-a^2)t^{\alpha-1}+\frac{a^2}{\pi}t^{2(\alpha-1)}
\;\;\;\;\;\;\; \mbox{for}\; t\gg 1 \mbox{ and } \; a\ll 1 \mbox{. }
\label{eqn:asasco}
\end{eqnarray}
Equation (\ref{eqn:asasco}) implies that the correlation function 
of the output signal obeys a power law $\sim t^{\alpha-1}$ with a small 
correction term. When the first term in (\ref{eqn:asasco}) is dominant, i.e. 
$a^2t^{\alpha-1}\ll 1$, the PSD of the output signal turns $f^{-\alpha}$.  
The condition $a^2t^{\alpha-1}\ll 1$ corresponds to $f \ll f_1$, where 
\begin{eqnarray}
f_1 \equiv (1/a^2)^{1/(1-\alpha)} \mbox{ . } 
\label{eqn:condi_f1}
\end{eqnarray}
The frequency $f_1$ can be very large
for small $a$ values because $1-\alpha>0$.
The power law spectrum $f^{-\alpha}$ is then observed 
in a wide range of frequencies.

For $a\gg 1$, an asymptotic expansion of (\ref{eqn:secfortw}) yields  
\begin{eqnarray}
\bar{y}=-1+\frac{2}{\sqrt{2\pi}}\exp\left[-\frac{a^2}{2}\right]
\cdot\bigm\{\frac{1}{a}+O(\frac{1}{a^3})\bigm\} \mbox{ . }
\label{eqn:ylargea}
\end{eqnarray}
We can calculate $T$ as follows. 
\begin{eqnarray}
T&=&2\int_{a}^{\infty}\!\!\!\int_{a}^{\infty}
\frac{dxdy}{A_{0}} 
\exp\left[-\frac{x^{2}-2cxy+y^{2}}{B}\right] \nonumber \\
%&=&2\int_{a}^{\infty}dy\int_{a-cy}^{\infty}
%\frac{dX}{A_{0}} \exp[-\frac{X^{2}+(1-c^2)y^{2}}{B}] \nonumber \\
&=&2\frac{1}{A_0}\int_{a^2}^{\infty}dY
\frac{1}{2Y^{\frac{1}{2}}}\exp\left[-\frac{1-c^2}{B}Y\right]
\int_{a-c\sqrt{Y}}^{\infty}dX\exp\left[-\frac{X^2}{B}\right]
\mbox{ . }
\end{eqnarray}
Integrating by parts with respect to the variable $Y$, 
we obtain
\begin{eqnarray}
\label{eqn:asggat}
T&=&E+F+G \mbox{ , } \\
E&\equiv&\frac{B}{A_0(1-c^2)}\frac{1}{a}
\exp\left[-\frac{1-c^2}{B}a^2\right]
\int_{(1-c)a}^{\infty}dX
\exp\left[-\frac{X^2}{B}\right] \mbox{ , } \nonumber \\
F&\equiv&\frac{B}{A_0(1-c^2)}\int_{a^2}^{\infty}dY
\frac{-1}{2Y^{\frac{3}{2}}}\exp\left[-\frac{1-c^2}{B}Y\right]
\int_{a-c\sqrt{Y}}^{\infty}dX\exp\left[-\frac{X^2}{B}\right] 
\mbox{ , } \nonumber \\
G&\equiv&\frac{B}{A_0(1-c^2)}\int_{a^2}^{\infty}dY
\frac{c}{2Y}\exp\left[-\frac{1-c^2}{B}Y\right]
\exp\left[-\frac{(a-c\sqrt{Y})^2}{B}\right] \mbox{ . } \nonumber 
\end{eqnarray}
The second term $F$ in (\ref{eqn:asggat}) can be neglected because 
$a \gg 1$. 
The asymptotic expansion of $E$ is obtained by substituting 
$X^2=x$ and 
integrating by parts as 
\begin{eqnarray}
E=\frac{B^2}{2A_0(1-c^2)(1-c)a^2}
\exp\left[-\frac{2a^2(1-c)}{B}\right]
+O\left(\frac{1}{a^4}\exp\left[-\frac{2a^2(1-c)}{B}\right]\right) \mbox{ . }
\label{eqn:tfirst}
\end{eqnarray}
In a similar manner 
G is calculated, 
by substituting $y=(\sqrt{Y}-ca)^2$ and 
integrating by parts, as 
\begin{eqnarray}
G=\frac{cB^2}{2A_0(1-c^2)(1-c)a^2}
\exp\left[-\frac{2a^2(1-c)}{B}\right]
+O\left(\frac{1}{a^4}\exp\left[-\frac{2a^2(1-c)}{B}\right]\right) \mbox{ . }
\label{eqn:tthird}
\end{eqnarray}
Equations (\ref{eqn:tfirst}) and (\ref{eqn:tthird}) yield 
\begin{eqnarray}
T=\frac{(1+c)^2}{\pi a^2 \sqrt{1-c^2}}\exp\left[-\frac{a^2}{1+c}\right]
+O\left(\frac{1}{a^4}\exp\left[-\frac{a^2}{1+c}\right]\right) \mbox{ . }
\label{eqn:tlarga}
\end{eqnarray}
Substituting  
(\ref{eqn:pasy}), (\ref{eqn:ylargea}), (\ref{eqn:tlarga}) and 
$c\equiv t^{\alpha-1}$ 
into (\ref{eqn:realascor}), we obtain 
\begin{eqnarray}
R_y(t)&\sim&\frac{2}{\pi a^2}\left[\frac{(1+c)^2}{\sqrt{1-c^2}}
\exp\bigm[-\frac{a^2}{1+c}\bigm]-\exp[-a^2]\right]
\;\;\mbox{ (for $a^2\gg1$)} \nonumber \\
&\sim& \frac{2}{\pi a^2}\left[(1+2t^{\alpha-1}+\frac{3}{2}
t^{2(\alpha-1)})\exp[-a^2(1-t^{\alpha-1}+t^{2(\alpha-1)}]
-\exp[-a^2]\right]\;\;\mbox{ (for $c\ll 1$).} \nonumber \\
&&
\label{eqn:asabco}
\end{eqnarray}
When $a^2t^{\alpha-1} \ll 1$, eq. (\ref{eqn:asabco}) reduces to 
\begin{eqnarray}
R_y(t)\sim \frac{2}{\pi}t^{\alpha-1}\exp[-a^2] \mbox{ . }
\end{eqnarray}
%Equation (\ref{eqn:asabco}) represents that the correlation function 
%of the output signal is $\sim t^{\alpha-1}e^{a^2}$ for 
%$a^2t^{\alpha-1} \ll 1$ i.e. $f\ll (1/a^2)^{1/(1-\alpha)}$. 
The above result leads to a $1/f^{\alpha}$ spectrum for the output signal 
in a low frequency range: $f\ll f_1$, where $f_1$ is defined by 
(\ref{eqn:condi_f1}).  
In contrast to the case $a\ll 1$, however, $f_1$ is small for $a\gg1$. 
This means that the power law spectrum $1/f^{\alpha}$ 
cannot be observed unless the 
observation time $t$ is large enough to satisfy $a^2t^{\alpha-1}\ll 1$. When 
this condition is not satisfied, $R_y(t)$ includes correction terms which 
are $t^{2(\alpha-1)}+\cdots$. 
Therefore PSD of the output 
signal deviates from the power law as 
 $A/f^{\alpha}+B/f^{2\alpha-1}+\cdots$. 
Since $\alpha > 2\alpha -1$, the second term makes the exponent of the PSD 
decrease. As the value of $a$ or $f$ increases, the higher order terms 
become more important, namely, the exponent tends to decrease. 
Figure.~\ref{fig:4} indicates this trend: The slope of the PSD turns 
flatter as $a$ or $f$ increases. 

\section{Construction of $1/f$ Noise from $1/f^2$ Noise} 
\setcounter{equation}{0}

 We present a numerical example to generate a $1/f$ noise 
from  $1/f^2$ noises using the dichotomous transformation 
introduced in \S 2. The procedure is as follows: 
%The idea is illustrated in Fig. 1: 
1) Prepare an ensemble of independent Gaussian $1/f^{\alpha}$ noises. 
2) Transform these noises to dichotomous ones via 
eq. (\ref{eqn:dichotomous}). As described in \S 2, 
PSDs of the transformed noises are proportional to 
$1/f^{(\alpha+1)/2}$. 
3) Average over a bunch of these dichotomous noises to re-Gaussianize, 
and prepare a number of such Gaussian $1/f^{(\alpha+1)/2}$ noises. 
4) Go back to 2). 

\begin{figure}[htbp]
\vspace{0.4cm}
	\begin{center}
	\includegraphics[width=0.92\linewidth]{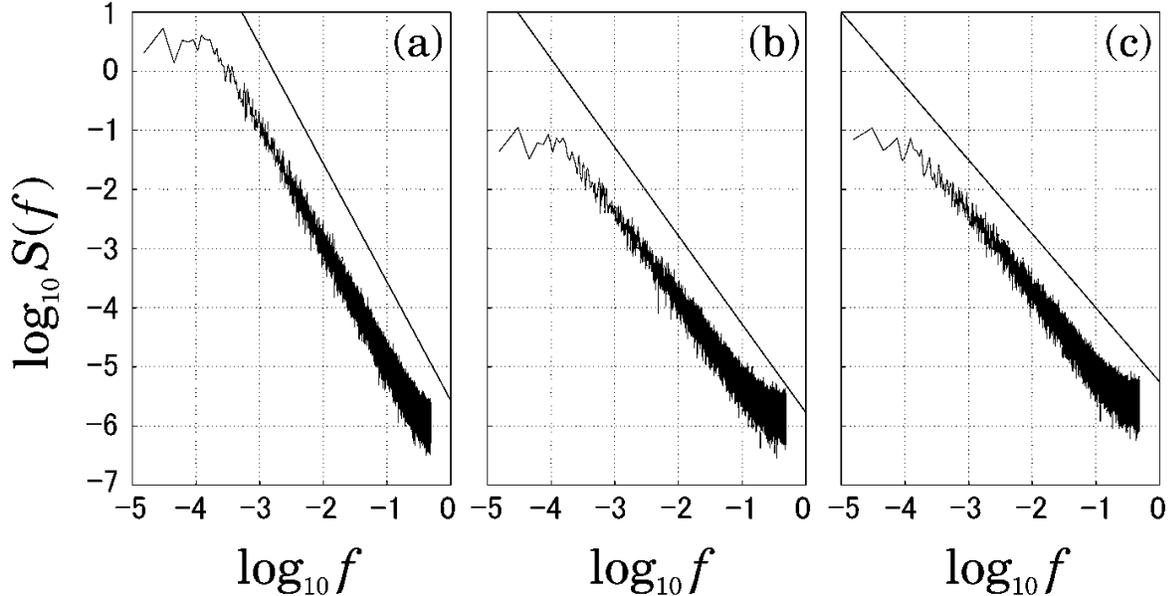}
	\end{center}
	\caption{PSD of the signal in each step in the procedure to achieve a 
$1/f$ noise. PSD is obtained by averaging over $10$ samples. 
The straight
lines represent $f^{-2}$ in (a), $f^{-1.5}$ in (b) and $f^{-1.25}$ in (c).
(a) PSD of a Lorentzian noise. (b) PSD of the noise transformed once.
(c) PSD of the noise transformed twice. To obtain a Gaussian input 
noise, five output noises by the previous transformation were averaged. 
}
\label{fig:6}
\end{figure}

We applied the above method to Lorentzian noises with a long 
relaxation time $\tau_1$, whose spectrum is 
$\sim [f^2+(1/\tau_1)^2]^{-1}$. The spectrum looks like $1/f^2$ 
when the observation time $t$ is not long enough: 
$t \ll \tau_1$ so that $f \gg 1/\tau_1$. The Lorentzian noises were 
generated by the following first order Markovian process 
\begin{eqnarray}
x(t+1)=\rho x(t)+n(t) \mbox{ ; }\;\;\; 
\rho=\exp\left[-\frac{1}{\tau_1}\right] \mbox{ , }
\label{eqn:lorentzian}
\end{eqnarray}
where $n(t)$ is white noise with zero mean. 
As $\rho$ approaches unity, the time series tends to a $1/f^2$ noise. 
A typical spectrum obtained from (\ref{eqn:lorentzian}) is shown in 
Fig.~\ref{fig:6}(a), where $\rho=1-0.001$, i.e. $\tau_1\sim 1000$ and 
$n(t)$ is a uniform random number in $[-0.5, 0.5]$. Starting with 
such Lorentzian noises, we obtained the first and the second 
transformations as Fig.~\ref{fig:6}(b) and Fig.~\ref{fig:6}(c), 
which are very close to $1/f^{1.5}$ and $1/f^{1.25}$ as was expected. 
It is rather amazing that only five independent output signals 
were averaged to obtain each {\it approximately} Gaussian noise. 
%%%%%%%%%%%%%%% fig. 5
%%%%%%%%%%%%%%%

There are several methods to generate $1/f^{\alpha}$ 
noise\cite{frieden94,dutta81,furukawa86,sano90} ; among them, McWhoter's 
model for $1/f$ noise is well known because 
the model corresponds to real physical systems, for 
example, noise due to 
surface traps of carriers in a semiconductor.\cite{dutta81} 
McWhoter's theory tells that if a large number of Lorentzian 
noises with various relaxation time $\tau$ are superposed with 
an appropriate weight: $w(\tau) \propto 1/\tau$ in a range 
$\tau_2 \leq \tau \leq \tau_1$, the spectrum 
becomes $1/f$ in the frequency range $1/\tau_1 \ll f \ll 1/\tau_2$.   
In contrast to McWhoter's theory, the present method requires 
only one kinds of Lorentzian noises specified by a single relaxation 
time $\tau_1$: $w(\tau)=\delta(\tau-\tau_1)$. Then the 
dichotomous transformation, if applied repeatedly, can lead to 
nearly $1/f$ noise in the range $f \gg 1/\tau_1$. 

In many cases, model systems exhibit $1/f$ noise only when the system 
parameters are tuned to some special values. For example, simulations 
of phonon number fluctuations based on a Fermi-Pasta-Ulam lattice 
resulted in a $1/f$ spectrum when the lattice size is $N=8$.\cite{fukamachi94,kawamura96} 
For larger $N$, however, the spectrum tended to Lorentzian 
$\sim (f^2+f_0^2)^{-1}$, although the time scale $\tau_1 \equiv 1/f_0$ 
became larger as $N$ increases.\cite{kawamura96} 
In contrast, tuning is unnecessary in the 
present method of generating $1/f$ noise, which is an essential point 
of the method. 

\section{Summary and Discussions} 
\setcounter{equation}{0}

We have analyzed the amplitude truncation of $1/f^{\alpha}$ noises 
(eq. (1.1)). Although $1/f^{\alpha}$ noises with 
$\alpha$ between $0$ and $2$ are commonly called 
`$1/f$ noise', the transformation property is different 
depending on whether $\alpha$ is larger than unity or not. 

For $\alpha > 1$, the output PSD under the symmetrical truncation 
obeys the power law with a smaller exponent $\beta=(\alpha+1)/2$ 
(which is still larger than unity) in a low frequency range 
($f_1 \ll $)$f\ll f_{\rm c}$. The corner frequency $f_{\rm c}$ depends on $\alpha$ 
and the truncation level $a$ as $f_{\rm c} \propto a^{-2/(\alpha -1)}$. 
Let us assume that the time scale of the system, $\tau_1=1/f_1$, is 
long enough so that the interval $[f_1$, $f_{\rm c}]$ is sufficiently wide. 
Then the filtering effect in the measurement makes it possible to 
observe a power law spectrum with the exponent much smaller than $2$, 
even if the original signal is close to Lorentzian: $\alpha \sim 2$. 
As was shown by the numerical simulation, if we start from an ensemble 
of Gaussian $1/f^2$ or Lorentzian signals and apply the symmetrical 
truncation and re-Gaussianization procedure repeatedly, the output 
signal converges to $1/f$ noise. 

When $\alpha\leq 1$, in contrast, the symmetrical truncation 
leads to a power law spectrum with the same exponent $\alpha$. 
(A proof for $\alpha=1$ has been given 
only for the dichotomous transformation.\cite{ishioka00}) This means that 
the exponent less than unity can never be reached by the symmetrical 
truncation as far as we start from the exponent larger than unity. 
On the other hand, when the asymmetrical amplitude truncation is 
applied to $1/f^{\alpha}$ signal with $\alpha$ less than unity, 
the output PSD deviates from the power law. The correction terms 
make the slope of the spectrum flatter as $f$ increases. However, 
as far as the results of the numerical experiments 
($\alpha=0.8$ in the present paper and $\alpha=1$ in ref. 5) are 
observed, the output PSD looks approximately as $1/f^{\alpha}$-like 
with an exponent smaller than that of the input signal. 

Amplitude truncation may generally occurs in systems composed of 
threshold elements, for example neural networks. It also occurs in 
the flow of packets in Internet systems 
where the overflow of the packets 
is deleted. 
Furthermore, measurement or analysis of signals often involves 
amplitude truncation. For example, a dichotomous transformation is 
used to analyze the (spatial) long-range correlation in DNA.\cite{stanly94} 
If a Lorentzian or $1/f^2$ noise is superposed on the original 
signal, filtering with such a truncation may easily leads to 
the observation of a $1/f^{\alpha}$ noise. 

Throughout the present paper, we have assumed that the time 
scale of the system is long enough so that the original data 
exhibits the power law spectrum in a wide range of frequencies. 
As mentioned already, to derive an extremely long time scale 
from any realistic model is a difficult problem. 
Once a long time scale has been assumed, however, the present 
study suggests a possible mechanism to generate $1/f^{\alpha}$ 
noises with the exponent smaller than that of the original signal.


\begin{thebibliography}{99}
\bibitem{rf:note}
Note that the output signal is generally non-Gaussian.
\end{thebibliography}
\end{document}